\documentclass[usenatbib,twocolumn,useAMS]{mn2e}
\usepackage{graphicx}    
\usepackage{latexsym}
\usepackage{dcolumn}
\usepackage{bm}
\input{epsf}
 
\def\prd{Phys. Rev. D} 
\def\apj{Astrophys. J.}         
\def\mnras{Mon. Not. R.  Astron. Soc.}

\newcommand{\bfig}{\noindent\begin{minipage}{3.48in}}
\newcommand{\efig}{\bigskip\end{minipage}}
\title[The IGM temperature and $y$]{The intergalactic medium temperature
and Compton y parameter}
\author[Pengjie Zhang, Ue-Li Pen \& Hy Trac]
{Pengjie Zhang\footnotemark, $^1$\thanks{E-mail:zhangpj@fnal.gov} Ue-Li Pen,
$^2$\thanks{E-mail:pen@cita.utoronto.ca} Hy Trac,
$^2$,$^3$\thanks{E-mail:trac@cita.utoronto.ca}\\ 
$1$NASA/Fermilab Astrophysics Group,
Fermi National Accelerator Laboratory,
Box 500,
Batavia, IL 60510-050\\
$2$Canadian Institute for Theoretical Astrophysics, University of
Toronto, Toronto, Canada, M5S 3H8\\
$3$Department of Astronomy \& Astrophysics, University of
Toronto, Toronto, Canada, M5S 3H8}

\begin{document}      
\maketitle

\begin{abstract}
The thermal Sunyaev Zeldovich (SZ) effect directly  probes the
thermal energy of the universe. Its precision modeling and 
future high accuracy measurements will provide a powerful way to
constrain the thermal history of the universe.  In this paper, we focus on the
precision modeling of the gas density weighted temperature $\bar{T}_g$
and the mean SZ Compton $y$ parameter. We run high
resolution adiabatic hydro simulations adopting the WMAP cosmology to
study the intergalactic medium (IGM) temperature and density distribution. To
quantify possible simulation limitations, we run $n=-1$,$-2$ self 
similar simulations.  Our
analytical model on  $\bar{T}_g$ is  based on energy  conservation and
matter clustering and has no free parameter. Combining both
simulations and analytical models thus provides the precision modeling
of $\bar{T}_g$ and $\bar{y}$.  We find that the simulated temperature
probability  distribution function and $\bar{T}_g$ shows good
convergence. For the WMAP cosmology, our highest resolution simulation
($1024^3$ cells, 100 
Mpc/h box size) reliably simulates $\bar{T}_g$ with better than $10\%$
accuracy for $z\ga 0.5$. Toward $z=0$, the simulation mass resolution
effect becomes stronger and causes the simulated $\bar{T}_g$ to be
slightly underestimated (At $z=0$, $\sim 20\%$ underestimated).  Since
$\bar{y}$ is mainly contributed by IGM at $z\ga 0.5$, such
simulation effect on $\bar{y}$ is no larger than $\sim
10\%$. Furthermore, our analytical model is capable of correcting
this artifact. It passes all tests of self similar simulations and 
WMAP simulations and is able to predict $\bar{T}_g$ and $\bar{y}$ to
several percent  accuracy. For low matter density $\Lambda$CDM
cosmology, the present $\bar{T}_g$ is $0.32
(\sigma_8/0.84)^{3.05-0.15\Omega_m}(\Omega_m/0.268)^{1.28-0.2\sigma_8}\
{\rm keV}$, which accounts for $10^{-8}$ of the critical cosmological density
and $0.024\%$ of the CMB energy. The mean $y$ parameter is $2.6\times 10^{-6}
(\sigma_8/0.84)^{4.1-2\Omega_m}(\Omega_m/0.268)^{1.28-0.2\sigma_8}$.
The current upper limit of $y<1.5\times 10^{-5}$ measured by FIRAS has
already ruled out combinations of high $\sigma_8\ga 1.1$ and high
$\Omega_m\ga 0.5$.  
\end{abstract}
\section{Introduction}
Ionized electrons with thermal motion can scatter CMB photons to
generate secondary 
 CMB temperature fluctuations known as the thermal  Sunyaev Zeldovich (SZ)
effect. Since all free electrons participate in the inverse Compton
scattering and contribute to the SZ effect, the SZ effect is an
unbiased probe of the thermal energy of the universe at $z\la 6$, for
which the universe is highly ionized. The thermal SZ effect is
sensitive to various physical processes like adiabatic  gravitational
heating, feedback, preheating and radiative cooling
\citep{daSilva01,Lin02,White02,Zhang03} which affect the thermal
energy of the baryons.  In addition, Compton cooling of first star
supernova  remnants \citep{Oh03}, cluster magnetic field
\citep{Zhang03b}, etc.  could further alter the thermal energy of the
universe to the level of $\ga 10\%$.  Therefore, the precision
measurement and interpretation  are of great importance to understand
the thermal history of  the universe.

Current CMB experiments such as
CBI\citep{Bond02,Mason03}, BIMA\citep{Dawson02} and ACBAR
\citep{Kuo02} marginally detected the SZ effect. Several
upcoming CMB experiments such as ACT, APEX, Planck, SPT  and SZA
are likely able to measure the SZ effect with $\sim 1\%$
accuracy in the next several years. In order to utilize the power of
such accurate experiments, the modeling of the SZ effect is required
to meet this $\sim 1\%$ accuracy. The first natural step for such
modeling is to robustly
understand the evolution of the baryon thermal energy in an
adiabatically evolving universe. It is not only
required to extract more complicated physics by comparing with
observations but also provides clues for the
modeling of these complicated physics.

Much effort has been devoted toward this goal, both analytically
\citep{Cole88,Makino93,Atrio-Barandela99,Komatsu99,Cooray00,Molnar00, 
Majumdar01,Zhang01,Komatsu02} and simulationally
\citep{daSilva00,Refregier00,Seljak01,Springel01,Zhang02}. Both
methods have limitations, which has not been quantified and
corrected to meet the precision of future observations.  Analytical
models of the SZ effect are  often {\it ad hoc} 
procedures. In the halo model, the cluster gas
pressure distribution is a free function of cluster mass and
redshift. Though it can be calculated by 
various assumptions such as hydrostatic equilibrium, its uncertainty
is hard to quantify.  In the continuum field model \citep{Zhang01},
the gas temperature is determined by the gravitational potential,
whose zero point is determined somewhat arbitrarily. So, analytical
models must be tested and  calibrated 
against simulations. For instance, \citet{Refregier02} has tested the
halo model against simulations and found good agreement.  However, the
conclusions drawn from these comparisons should be viewed with some
caution.  Numerical simulations are known to have 
artifacts, stemming from limited resolution and finite volume.  The
impact of some of the artifacts have been investigated for the thermal
SZ effect \citep{Refregier02} and the kinetic SZ effect
\citep{Zhangpj03}. If not corrected, such artifacts would lead  to
biased calibrated analytical models.

The SZ mean temperature decrement, or equivalently, the mean SZ
Compton $y$ parameter, which corresponds to the density weighted gas mean
temperature  $\bar{T}_g$, are the lowest order SZ statistics. They are
also the easiest to simulate and  model. So, the precision
prediction of $\bar{T}_g$ and $\bar{y}$  stands as the first natural
step toward the  precision modeling of the SZ effect.  Their precision
modeling also provides clues  for the next low order SZ statistics
such as  the SZ power spectrum and the  
corresponding gas pressure power spectrum. In this paper, we present a
detailed study of the IGM density and temperature distribution from a
series of $\Lambda$CDM and self similar simulations. We further
test the continuum  model prediction of $\bar{T}_g$ and $y$ parameter against
simulations. Our goal is to  quantify and correct numerical
limitations and build  calibrated analytical model aimed at 
$1\%$ accuracy. We will
follow a similar procedure as in this paper to discuss the precision
modeling of the SZ power spectrum in a companion paper (Zhang, Pen \&
Trac, 2004, in preparation).

\section{The thermal Sunyaev-Zeldovich effect}
Free electrons scatter off CMB photons by their thermal motions and introduce
secondary CMB temperature fluctuations:
\begin{equation}
\Theta=-2 y S_T(\nu)=-2y \left[2-\frac{x/2}{\tanh(x/2)}\right],
\end{equation}
where $x\equiv h\nu/k_BT_{\rm CMB}$.  This is known as the thermal
Sunyaev-Zeldovich (SZ) effect \citep{Zeldovich69}. The Compton {\it y}
parameter is given by the integral of electron thermal pressure energy
along the line of sight 
\begin{equation}
\label{eqn:y}
y=\int \frac{n_e k_B T_e}{m_ec^2} \sigma_T a d\chi,
\end{equation}
where $\chi$ is the comoving distance  and $a$ is the scale
factor. The lowest order statistics of the SZ effect is the 
mean Compton $y$ parameter: 
\begin{equation}
\bar{y}=-2.37\times 10^{-4} \Omega_bh \int \frac{\bar{T}_g}{\rm
keV}a^{-2} d\tilde{\chi},
\end{equation}
where $\bar{T}_g\equiv \langle (1+\delta_g)T_g \rangle$ is the gas density
weighted mean temperature and $\tilde{\chi}\equiv \chi/(c/H_0)$ is
the dimensionless comoving distance while $H_0$ is the present Hubble
constant.  

The thermal energy of the universe only accounts for a tiny fraction
of the total energy of the universe. 
\begin{equation}
\Omega_{\rm TE}=1.08\times 10^{-8} \frac{\bar{T}_g}{0.3 {\rm
keV}}\frac{\Omega_b h^2}{0.02}\ .
\end{equation}
As a comparison, $\Omega_{\rm CMB}=2.48\times 10^{-5} h^{-2}$ and the
energy in other wavelength bands of light 
\begin{equation}
\Omega_{\rm EBL}=2.48\times 10^{-6} h^{-2} \frac{I_{\rm EBL}}{100\ {\rm nw\
m^{-2}\ sr^{-1}}}\ .
\end{equation}
\begin{figure}
\epsfxsize=9cm
\epsffile{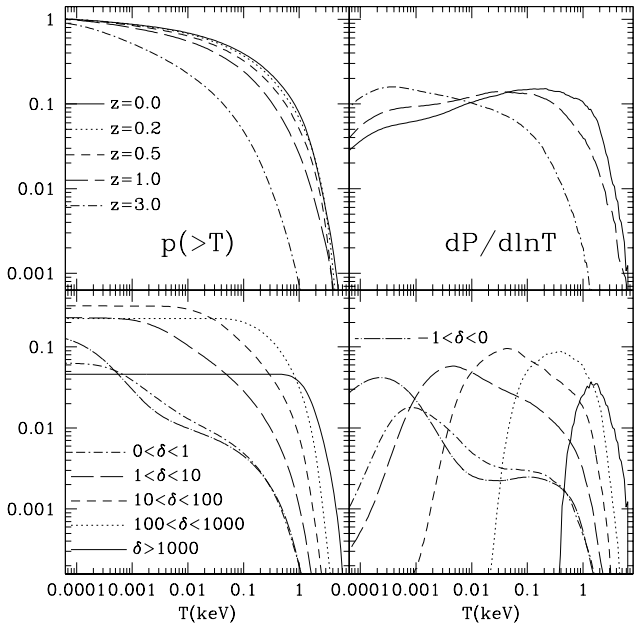}
\caption{The gas temperature distribution in the $1024^3$, $100$ Mpc/h
WMAP simulation. The top left panel shows the cumulative mass fraction
$p(>T)$ with
temperature higher than $T$. The bottom bottom panel shows the
conditional $p(>T)$
in different  overdensity regions at $z=0$.  The
right panels plot the corresponding ${\rm d}P/{\rm d}\ln T$. At
$z=0.0$, only $5\%$ gas is 
hotter than 1 keV. Gas temperature is tightly correlated with gas
density.  Nearly all virialized gas is hotter than $1$ keV while
effectively no gas with $\delta<10$ is hotter than $0.1$
keV. \label{fig:WMAPd}} 
\end{figure}

\section{Hydro simulations}
We ran cosmological hydrodynamical simulations using a new Eulerian 
cosmological hydro code  \citep{trac03a,trac03b}.  This Eulerian code
(hereafter TP) is based on the finite-volume, flux-conservative total variation
diminishing (TVD) scheme that provides high-order accuracy and high-resolution
capturing of shocks.  The hydrodynamics of the gas is
simulated by solving the Euler system of conservation equations for mass,
momentum, and energy on a fixed Cartesian grid.  The gravitational evolution of
the dark matter is simulated using a cloud-in-cell particle-mesh (PM) scheme
\citep{hockney88}.

The robustness of  the TP code has been tested by comparing the evolution of
the dark matter and gas density power spectra from the simulations with the
fitting formula of \citet{Smith03}.  We also performed 
a code comparison by running
the same initial conditions using the MMH code \citep{Pen98}, which
combines the 
shock capturing abilities of Eulerian schemes with the high dynamic range in
density achieved by Lagrangian schemes.  Power spectra are computed using FFTs.
We find good agreement at all relevant scales and redshifts for both
comparisons.

Eulerian schemes are ideal for simulating the
evolution of the IGM to model the thermal and kinetic SZ effects and
the Lyman alpha  
forest, because of their high speed, superior mass resolution,
shock-capturing abilities.  Furthermore, Eulerian algorithms
are computationally very fast and memory friendly, allowing one to optimally
use available computational resources.

\begin{figure}
\epsfxsize=9cm
\epsffile{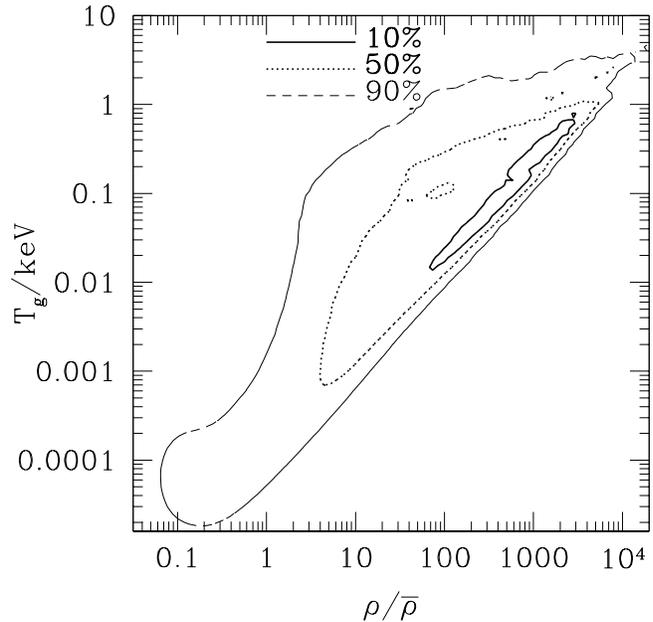}
\caption{The contour of gas (log)temperature and (log)density in the $1024^3$
WMAP simulation. Gas temperature strongly correlates with gas density
with a scaling relation $T\propto \rho$, as found in previous works (e.g. \citet{Kang94,Dave01}). \label{fig:WMAPcontour}}
\end{figure}
\begin{figure}
\epsfxsize=9cm
\epsffile{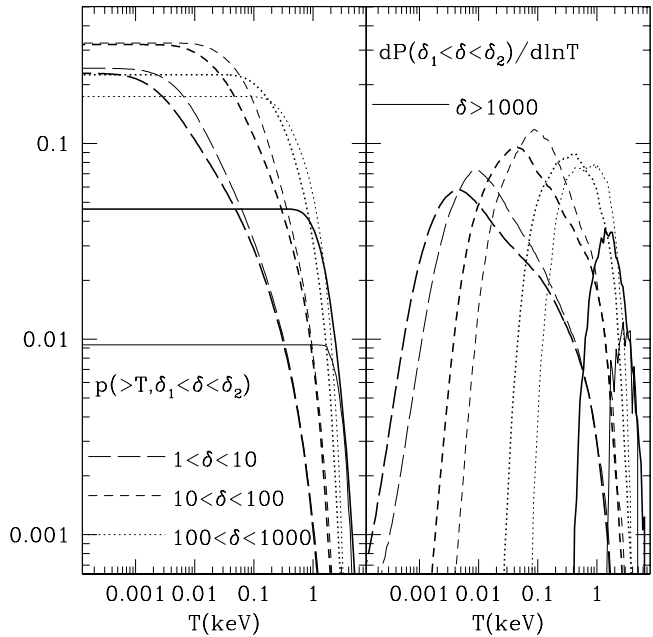}
\caption{The comparison of the conditional  temperature distribution
function $p(>T,\delta_1<\delta<\delta_2)$ between WMAP $1024^3$ (thick
lines) and $512^3$ (thin lines) simulations. For clarity, we only
show $\delta>1000$ (solid lines), $1000>\delta>100$ (dot lines),
$100>\delta>10$ (short dash lines), $10>\delta>1$ (long dash
lines). Though the overall $p(>T)$ agrees very well for two
simulations (Fig. \ref{fig:WMAPcom1}), the conditional
$p(>T,\delta_1<\delta<\delta_2)$ does not 
converge yet and reflects the effect of simulation
resolution. $\delta\ga 1000$ corresponds to the inner regions of
clusters and groups. For these virialized objects, their temperature
scales with respect to their mass $M$ as $M^{2/3}$. For clusters with
virial radius around $1$ Mpc/h, $T\sim 1$ keV. Higher
resolution simulations are able to resolve more low mass halos, which
are smaller and have lower $T$ and thus result in a higher
$p(>T,\delta>1000)$ at $T\la 
1$ keV.  The limited simulation resolution introduces artifacts
into the $T$-$\rho$ correlation. Though its effect to the gas density
weighted temperature is minor (Fig. \ref{fig:WMAPcom1}), it would
affect the simulated pressure power spectrum
significantly. \label{fig:WMAPcom2}}  
\end{figure}
\begin{figure}
\epsfxsize=9cm
\epsffile{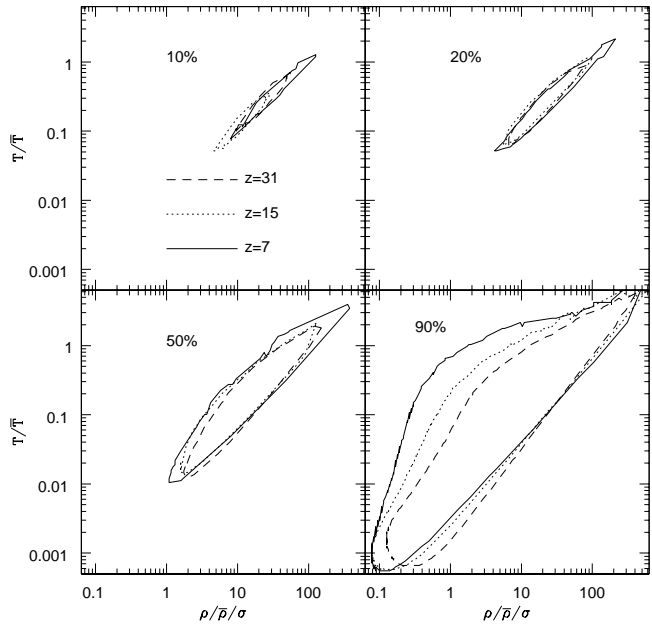}
\caption{The contour of gas (log)temperature and (log)density in the
$n=-1$ self  similar simulation.  The simulations show good convergence in
most $T$-$\rho$ regions. Higher resolution is required to simulate both
low $T$, low $\rho$ regions and high $T$, high $\rho$ regions. But low
$T$, low $\rho$ regions have little contribution to mean gas density
weighted temperature. 
\label{fig:tdn12}}
\end{figure}
We ran a $1024^3$ cells, $100$ Mpc/h box size  simulation with the
best fit WMAP-alone  cosmology $\Omega_m=0.268$,  
$\Omega_\Lambda=0.752$, $\Omega_b=0.044$, $h=0.71$, and
$\sigma_8=0.84$ \citep{Spergel03}. The
ratio of dark matter particles to fluid elements is 1:8. We achieve a spacial
resolution of $\Delta x\simeq100$ kpc/$h$ and a dark matter particle mass
resolution of $\Delta m\simeq5.6\times10^8\ M_\odot$.  The initial
conditions are 
generated by sampling from an initial power spectrum computed using CMBFAST
\citep{CMBFAST}.  This simulation is started at a redshift of $z=100$ and
evolved down to $z=0$, with data outputs at $z=$ 3, 1, 0.5, 0.2 and
1. This simulation takes approximately  $700$ time-steps to evolve from
$z=100$ down to  $z=0$.  On a GS320 Compaq Alpha server with 32 cpus
and total theoretical peak speed of 32 Gflops, the run {\bf requires
40 G memeory} and takes 
approximately two days. Simulations are limited by both the box size,
which causes the absence of large scale density fluctuation at scales
larger than half box size, 
and resolution, which results in the failure to resolve small scale
structures. Self similar simulations are ideal to test and
quantify such simulation limitations since different redshifts
directly corresponds to different resolution and box size. We ran one
$n=-2$ and one $n=-1$ ($\Omega_m=1$)  self similar 
simulation with $512^3$ 
cells and  the same amount of dark matter particles.  In these
simulations, $\Omega_b$ is set to be $0.044/0.268$ to mimic the
$\Omega_b/\Omega_m$ ratio of the WMAP cosmology.  The initial
fluctuation is normalized such that, when linearly extrapolated to
$z=0$, the correlation length is half the simulation box size. We also run a
$512^3$ cells, 100 Mpc/h WMAP simulation with the identical initial
condition as the $1024^3$ for a direct comparison. The moving frame of
the TP code is not turned on in these runs.  We will check its effect
in the future. 

We show the temperature distribution function, namely the mass
fraction of gas hotter than $T$, $p(>T)$ of the $1024^3$ WMAP
simulation in Fig. \ref{fig:WMAPd}. At $z=0.0$, only $5\%$ gas are hotter
than $1$ keV and  $\sim 40\%$ gas are hotter than $0.1$ keV. These
fraction drops to $3\%$ and $24\%$ at $z=1.0$. Gas temperature shows a
strong positive correlation with its density. At $z=0.0$, nearly all
gas with $\delta\ga 1000$ are hotter
than $1$ keV. For $100<\delta<1000$, almost all gas are hotter than 0.1
keV. Nearly no gas hotter than $0.1$ keV lies in $\delta<10$
region. It is interesting to see how much gas lies in virialized
halos. Virialized gas should have an overdensity larger than that of
the gas overdensity at the virial radius. For  an
isothermal density profile $\rho\propto r^{-2}$, this states
$\delta\geq \delta(r_{\rm vir})= 1/3 \Delta_c/\Omega_m\sim
100$. $\Delta_c\sim 100$ for WMAP 
\citep{Eke96} is the mean matter density in a virialized halo with
respect to the critical density.   This factor $1/3$ does not change
much for other profiles such as NFW and thus we omit its
variation. Then the bottom panel of Fig. \ref{fig:WMAPd} implies that only
$\sim 28\%$ gas resides in virialized halos. At $z=1.0$, this fraction
drops to $\sim 19\%$. 

In our simulation, a large fraction of gas is colder than $\sim 10^4
$K. At $z=0$, this fraction is $\sim 10\%$ and at $z=3$, this fraction
reaches $\sim 50\%$. In reality, part of these gas may condense into stars
or  interstellar medium. Part of them may be photoionization-heated
to above $10^4$K. Our simulation does not include any photoionization,
radiative cooling, etc., so the prediction about these gas is not
reliable and is hard to compare with other works (e.g. \citet{Kang94,Dave01}). But the contribution of such gas to the SZ effect is
negligible due to their low temperature, so we omit the complexity
caused by such gas. 

\begin{figure}
\epsfxsize=9cm
\epsffile{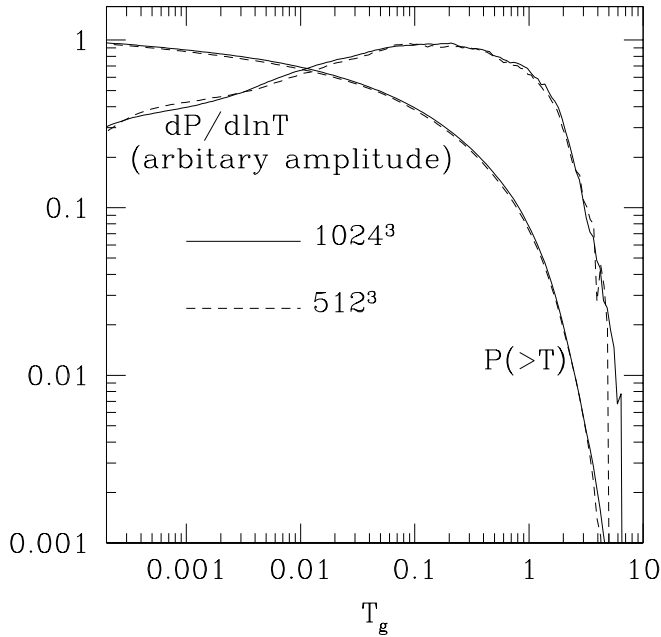}
\caption{The comparison of the overall temperature distribution
function $p(>T)$ between the WMAP $1024^3$ and the $512^3$
simulations. Two simulations agree well. $p(>T)$ is directly related
to the mean gas mass weighted temperature $\bar{T}_g$ through the
relation $\bar{T}_g\equiv \int T{\rm d}p$.  The good convergence of
$p(>T)$ implies that the simulation effects to $\bar{T}_g$ is minor,
though we will still  quantify them in \S
\ref{sec:model}. \label{fig:WMAPcom1}} 
\end{figure}

The strong correlation of $T$ and $\rho$ observed in $p(>T,>\rho)$ is
clearly shown in the  $T$-$\rho$ contour of our $1024^3$ simulation
(Fig. \ref{fig:WMAPcontour}). $T\propto \rho$ holds in a large
$T$-$\rho$ regions, as found in previous works (e.g. \citet{Kang94,Dave01}).

Simulations are both mass and spacial resolution limited. So the above
results may be strongly resolution dependent. Both the Press-Schechter
formalism \citep{Press74} and the Jenkins fitting formula \citep{Jenkins01}
imply the existence of numerous small halos. The failure to resolve
these small halos will result in  an underestimation of the fraction of
virialized gas and biased $T$-$\rho$ relation.  To estimate the
resolution effect, we compare between WMAP simulations and between
self similar simulations.

In low density regions, the density field is well 
resolved. But such regions generally have high Mach number and thus
shocks are relatively poorly resolved. Since gas thermal energy is
generated by shock heating, in low density regions, temperature field
is likely poorly resolved. In high density regions, it is the opposite
case.  Fig. \ref{fig:WMAPcom2} shows that both fields are indeed
significantly  affected by resolutions.  The fraction of $\delta>1000$
gas increases from $\sim 1\%$ in the $512^3$ simulation to $\sim 5\%$
at $1024^3$. The $1024^3$ simulation is able to resolve smaller halos, which
have lower temperature and results in an increase in $dp(>T,
\delta>1000)/d\ln T$ at $T\la 3$ keV.  The fraction of  virialized gas
($\delta\ga 100$) increases from $\sim 18\%$ to $\sim 28\%$. For
$\delta<10$, the conditional $p(>T)$ agrees well at the high $T$ tail
where Mach numbers are
low and diverges at the low $T$ tail where Mach numbers are high. So, for
$\delta<10$, shock capture is the dominant resolution factor. In summary, the IGM density and temperature distribution in simulations is
mainly limited by density resolution in high density regions and shock
capturing ability in low density regions. We thus expect that higher
resolution simulations will result in larger dispersion in the gas
$\rho$-$T$ phase space distribution, and thus larger pressure
dispersion.  This conclusion is further confirmed in the $T$-$\rho$
contours of the  probability   distribution for our self
similar  simulation results (Fig. \ref{fig:tdn12}).  Such resolution
effect has only minor effect on $\bar{T}$, but its effect on the
pressure power spectrum may be important. We  will study this issue
in a companion paper (Zhang, Pen \& Trac, 2004, in  preparation).

{\bf 
Our simulations suggest that (1) only a small fraction of
gas are virialized and (2) a strong correlation exists between gas
temperature and  density. Both apparently contradict with the halo
model. The halo model assumes that all gas resides in virialized
halos. This assumption seems to contradict with (1), as noticed by
\citet{Refregier02}.  The simplest
version of the halo model assumes gas to be isothermal 
with temperature equal to its virial temperature. Since halos roughly
have the same density, the predicted $T$-$\rho$ correlation should be
very weak.  The weak dependence of halo concentration number on halo
mass can not alter this straight prediction. More complicated
intracluster gas models such as the {\it universal gas profile}
\citep{Komatsu01}   
predict a weak variation of gas temperature from the core to the
virial radius. But such variation is still too weak to explain the
simulated $T\propto \rho$ relation. 

But simulations are resolution limited and some artifacts can be
clearly seen in Fig. 3. It is likely that the part of the
contradictions discussed 
above are caused by the limited simulation
resolution. The limited simulation resolution 
effectively smooths the density field and may cause an underestimation
of the fraction of virialized gas. The spacial resolution ($\sim 0.1$
Mpc/h for our largest simulation) instead of
the mass resolution is the main limiting factor to resolve halos. If
our simulation only  resolves halos  with virial radius $r_{\rm
vir}\geq r_{\rm min}\sim 0.5 $ Mpc/h, following the Press-Schechter formalism
\citep{Press74}, such halos contain  $f_h=1-{\rm
Erf}[-\nu/\sqrt{2}]\sim 30\%$ of the total mass at $z=0$. Here,
$\nu\equiv \delta_c/\sigma(m_{\rm min})$, where $\delta_c\sim 
1.686$ is the density threshold, $\sigma(m)$ is the density
fluctuation in a sphere with mean mass $m$ and $m_{\rm min}$ is the
mass of halos with $r_{\rm vir}=r_{\rm min}$. This mass fraction is
consistent with our simulation results. But this predicted mass
fraction is very sensitive to the $r_{\rm min}$ assumed.  For example,
if $r_{\rm min}\sim 0.3$ Mpc/h, the predicted virialized gas fraction
would be $f_h\sim 50\%$ and thus contradicts with our simulation
result. One can also compare the prediction of $p(>T)$ of halo model
with our simulation. The gas temperature $T$ of a halo mass $m$ can be
estimated by $T=T_8 (m/m_8)^{2/3}$, where at $z=0$, $T_8\sim
5\Omega_m\simeq 1.3$ keV \citep{Pen98b} and $m_8$ is the mean mass
contained in a 
sphere with radius $8$ Mpc/h. Thus the halo model predicts $p(\ga 1.3
{\rm keV})=f_h(m\geq m_8)\sim 4\%$, and $p(\ga 0.1 {\rm keV})\simeq
f_h(m\geq 0.02m_8)\sim 50\%$. These predictions are roughly consistent with our
simulation result. Thus, a low fraction of virialized gas may be
caused by simulation resolution.

The effect of resolution to the $\rho-T$ relation can be estimated as
follow.\footnote{The discussion presented here is based on the
comments of the anonymous referee.}. The density profile of each halo
can be approximated as  $\delta=\delta_{\rm vir} (r/r_{\rm vir}\sim 100
(r/r_{\rm vir})$.  The smoothing caused by limited resolution convolves
this density field with a window function with radius $R_f$. For halos
with $r_{\rm vir}\la R_f$, the smoothed density $\delta^f$ will be
$\delta^f \sim 100 (r_{\rm vir}/R_f)^{2}\propto T$. Here, we have
utilized that $r_{\rm vir}\propto M^{1/3}\propto T^{1/2}$.   For WMAP
cosmology at $z=0$, this implies that $\delta\sim 100(1{\rm Mpc/h}/R_f)^2
(T/{\rm keV})$.  This relation is surprisingly consistent with our
simulation, if taking $R_f\simeq 0.2$ Mpc/h.  So the $\rho-T$ relation
may also be caused by artifacts of simulations and has to be used in
caution.

On the other hand, the simulation resolution is hard to explain all
aspects of simulation results. The $\rho$-$T$ relation shows a robust
convergence in the self similar simulations (e.g. 20\%, 50\% $\rho$-$T$
contour in Fig. 4) and only changes very slowly as the
non-linear length scale  changes by a factor of 4. This suggests that
it is unlikely that the  observed $\rho$-$T$ relation is purely caused
by simulation artifacts, otherwise  the $\rho$-$T$
relation would  change significantly with resolution. 

These issues deserves further 
investigations. But more careful comparison requires more realistic halo
mass function than the Press-Schechter formalism and a series of
simulations to have fair sample of $\sim $ keV halos and thus beyonds
the scope of this paper.}

\begin{figure}
\epsfxsize=9cm
\epsffile{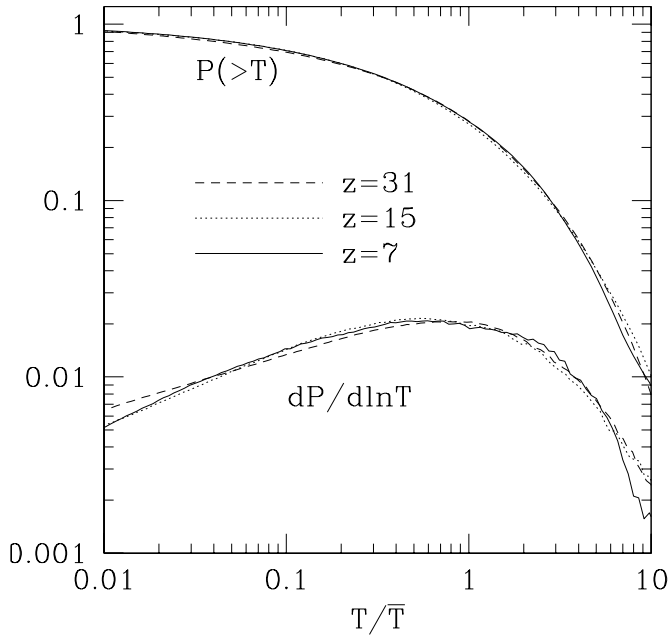}
\caption{The overall temperature distribution $p(>T)$ for $n=-1$ self similar
simulation. For the temperature field, the simulation effect is minor
except for very low or high temperature regions. Since the gas density
weighted temperature $\bar{T}_g=\int T {\rm d}p$,  $\bar{T}_g$ is well simulated.\label{fig:tdn11}}
\end{figure}

But despite these significant resolution effects on both density and
temperature fields, the simulated overall $p(>T)$ shows a good convergence 
except at very low or  high temperature ranges
(Fig. \ref{fig:WMAPcom1} \& \ref{fig:tdn11}). Since the energy
conservation guarantees the total amount of kinetic and thermal energy
to be well simulated, the good agreement of $p(>T)$ implies that the
conversion efficiency from kinetic energy to thermal energy is well
simulated too. Since  $\bar{T}_g\equiv \int Tdp$,  the effect of
simulation limitations  to $\bar{T}_g$ should be 
minor, as we will quantify in the next section. 

The simulated $\bar{T}_g(z)$ would be less resolution
dependent and  robust. We will 
develop our model for $\bar{T}_g(z)$ in the next section, test it
against simulations, quantify simulation artifacts and provide a
precision model of $\bar{T}_g(z)$. As we will show in \S
\ref{sec:model}, the simulation effect on  $\bar{T}_g$ is only
non-negligible at $z\la 0.5$ and this effect is no bigger than  $\sim
20\%$, even at $z=0$.

\section{The continuum field model}
\label{sec:model}
In a gravitational heating scenario, the gas temperature is determined
by the 
gravitational potential $\Phi$.  The pressure depends on the
thermalized fraction of the total kinetic energy.  The translational
kinetic energy is thermalized from the
energy released when particles shell cross.  A model of the
thermalized energy is thus given by the difference in energy between 
two particles separated by a non-linear scale in Lagrangian space,
which is the distance at which they can be expected to have shell
crossed.  The exact procedure amounts to solving the non-linear
evolution equations directly.  But we can treat the effect
statistically in a linear fashion.  In the initial linear evolution, the
gravitational potential remains constant.  After virialization, the
gravitational energy at a fixed location remains almost
constant.  In an Eulerian description, we can describe the energy of
particles at a final virialized location as the energy released as a
particle travels from its initial position to the final virialized
location.  We can then relate the gas temperature to
$\Phi$ through the viral theorem: 
\begin{equation}
k_B T_g({\bf x})=\frac{1}{6}\frac{4m_H}{3+5X}\left[\Phi({\bf x})-\bar{\Phi}({\bf x})\right].
\end{equation}
Since  the initial position is not exactly known, we take a
spherical average over the non-linear scale to average over all
possible initial locations and obtain the mean initial potential
\begin{equation}
\bar{\Phi}({\bf x})=\int \Phi({\bf x}^{'}) W_e({\bf x-x}^{'}) d^3x^{'}.
\end{equation}
For a detailed 
explanation, refer to \citet{Zhang01}. Then, the averaged gas density weighted temperature  
\begin{equation} 
\label{eqn:meanT}
\bar{T}_{g}=0.016 \Omega_m {\rm keV} (1+z)\int_0^{\infty}
\Delta^2_{\rm dg}(k)f_e(k) \frac{dk}{k}.
\end{equation}
Here, $f_e(k)\equiv [1-W_e(k)]/k^2$ ($k$ is in unit of
Mpc/h). $\Delta^2_{\rm dg}$ is the dark matter-gas cross correlation
power spectrum. 

\begin{figure}
\epsfxsize=9cm
\epsffile{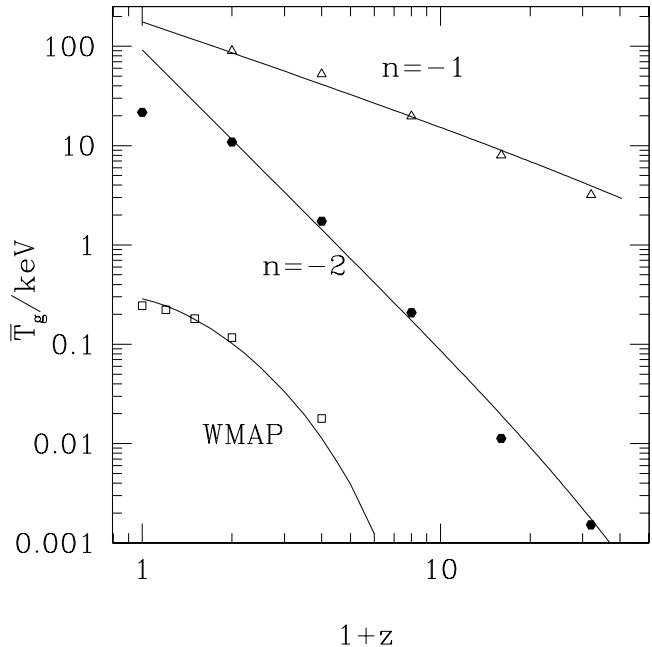}
\caption{The density weighted gas temperature $\bar{T}_g$. Data points
are the results of our self similar simulations ($512^3$), WMAP
simulation ($1024^3$, 100 Mpc/h) and solid
lines are our model predictions. Our model predicts a self similar
scaling relation  $\bar{T}_g(z)\propto (1+z)^{(n-1)/(n+3)}$ for self similar
simulations. Our predictions agree with 
simulations very well at most redshifts. The breaking of the self
similar relation at low redshifts suggests the limitation of
simulations. We will show that the
discrepancy of the WMAP simulation at low redshifts is caused
by simulation resolution in fig. \ref{fig:wmapTz} while the
discrepancy of the self similar simualtions is caused by limited box
size.\label{fig:Tz}} 
\end{figure}

In our model, $W_e(k)$ is a free function, but its asymptotic behavior
toward $k=0$ is fixed by the requirement that $T_g$ follows the
density field at large scales, or equivalently, the $T_g$ bias with
respect to the underlying density field is a constant at large scales. Its behavior at
small scales is hard to determine from first principles. But since at scales
smaller than smoothing  scale, $W_e(k)\rightarrow 0$ and
$f_e(k)\rightarrow k^{-2}$,  the exact behavior of
$W_e(k)$ at large  $k$ is not very important. Based on these
considerations, a natural  choice of $W_e(k)$ is a 
Gaussian function $W_e(k)=\exp(-k^2r_e^2)$. For this function, when
$k\rightarrow 0$, $f_e(k)\rightarrow r_e^2$, so the temperature bias
with respect to the density field is a constant. Since gas gains
thermal energy by shell crossing, which happens at the nonlinear
scales, we expect $r_e$ to be roughly equal to the 
density correlation length. So, the evolution of $r_e(z)$ follows that
of the density correlation length.  Essentially, the only free
parameter in our model is $r_e(z=0)$.  We will choose
$r_e=r_{\rm NL}$, where $r_{NL}$  is the non-linear scale 
set by $\xi_L(r_{\rm NL})=1$ with $\xi_L$ as the linear correlation
function. This choice of $r_e$ has to be tested against simulations
and this is the only parameter that requires calibration against
simulations. 

\begin{figure}
\epsfxsize=9cm
\epsffile{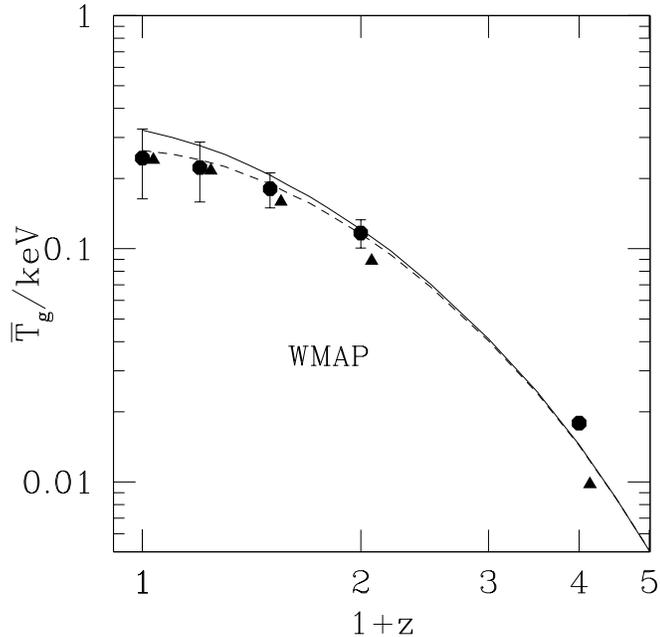}
\caption{The resolution effect of simulations to $\bar{T}_g$. The data
points with (2$\sigma$) error bars are the WMAP simulation result while the
triangle data points are our model prediction using the simulated
$\Delta^2_{\rm dg}$. For clarity, triangle data points are shifted
horizontally arbitrarily. The solid line is our model prediction assuming gas
perfectly follows dark matter and the dash line is the prediction
assuming a small scale cutoff in the gas density power spectrum. The
excellent agreement between two sets of data points supports the
validity of our model 
and implies the simulation limitations to be the cause of the apparent
discrepancy found in fig. \ref{fig:Tz}. The excellent reproduction of
the simulation results at low redshifts by the dash line implies that
the simulation  resolution  is the cause of the discrepancy. 
\label{fig:wmapTz}} 
\end{figure}
\begin{figure}
\epsfxsize=9cm
\epsffile{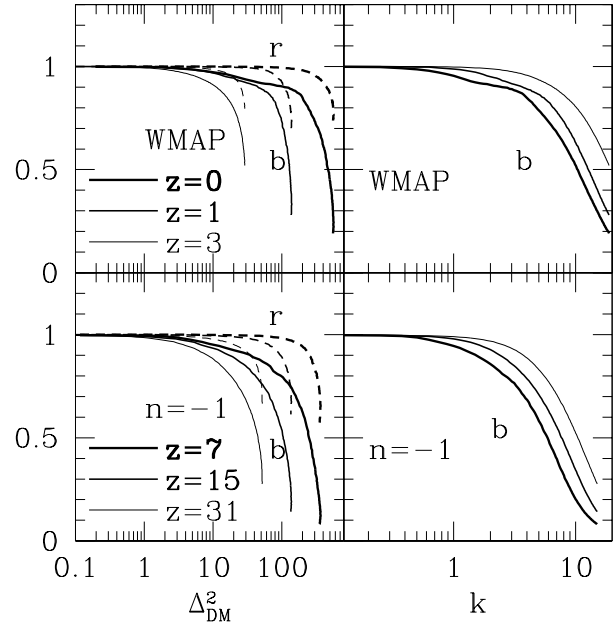}
\caption{The gas-dark matter relation. The dash lines are the gas-dark
matter cross correlation coefficients while the solid lines are the gas
biases versus dark matter at different redshifts.  At small nonlinear
scales, gas ceases to 
follow dark matter in simulations. But the 
redshift dependence of such behavior suggests that it is unphysical
and possibly caused by the simulation resolution effect. Lower
redshift simulation has better mass resolution (left panel), which
corresponds to better spacial resolution in unit of nonlinear
scale. But spacial resolution in unit of  absolute physical scale is
poorer  at lower $z$ (right panel). \label{fig:r}} 
\end{figure} 
For self-similar simulations, $\Delta^2_{\rm dg}(k)$
should scales as $f(k/k_{\rm NL})$ (We define $k_{\rm NL}$ as $k_{\rm
NL}\equiv 1/r_{\rm NL}$),  then Eq. \ref{eqn:meanT} naturally predicts 
\begin{equation}
\label{eqn:selfsimilarTz}
\bar{T}_g(z)=\bar{T}_G(z=0) (1+z)^{(n-1)/(n+3)}.
\end{equation}
Without feedback or cooling, the  gas should follow the  dark matter
distribution 
matter to very high overdensity and thus we expect
$\Delta^2_{\rm dg}(k)=\Delta_{\rm dm}^2(k)=\Delta^2_g(k)$.

 We calculate $\Delta^2_{\rm dm}(k)$ using the code of
\citet{Smith03}. We compared the predictions from our model with 
simulation results and found a good agreement (Fig. \ref{fig:Tz}). The
scaling relation  Eq. (\ref{eqn:selfsimilarTz}) with the right
amplitude is observed at high
redshifts and further confirms the validity of our model. 

At low redshifts, the scaling relation breaks down for the self similar
simulations. This is caused by the finite simulation box size.  Its
effect to  $\bar{T}_g$ corresponds to  a lower $k$ cutoff $k_{\rm cut}=2\pi/L$ in the integral of Eq. \ref{eqn:meanT}, where $L$ is the box
size. For our 
self similar simulations, the correlation length at $z=0$ is half
the box size and $k_{\rm NL}=2/L$. One has $k_{\rm cut}=\pi k_{\rm
NL}$. So, it is the  limited box  size that causes 
$\bar{T}_g$ in self similar simulations to lose power at low redshifts. 

The deviation between the predicted and simulated $\bar{T}_g$ is also observed
in the WMAP simulation. Such discrepancy increases toward low
redshifts and exceeds  $2\sigma$ level at $z=0$
(Fig. \ref{fig:wmapTz}), so it is hard to be  explained by sample
variance. If this discrepancy is caused by
simulation limitations, $\bar{T}_g$ calculated using the simulated WMAP
$\Delta^2_{\rm dg}(k)$ should agree with the simulated
$\bar{T}_g$. Indeed, the agreement is better than $5\%$ at low
redshifts (Fig. \ref{fig:wmapTz}). Thus we show that this
discrepancy can be naturally explained by the simulation limitations
and  thus our model works well to better than several percent at low
redshifts.

We further probe which simulation limitation causes this
discrepancy. For WMAP simulations, even at $z=0$, the nonlinear scale
is still   much smaller than the box size, so  the box size effect is
negligible.  Resolution effect causes $\Delta^2_{\rm dg}$ to lose power at
small scales and causes the simulated $\bar{T}_g$ to lose power. Since the
resolution of the hydro part of a hydro simulation is 
generally worse than that of the  N-body part, gas ceases to follow
dark matter below certain scale. Such deviation is a suitable measure
of simulation resolution and can be quantified. We  define the gas
bias $b_g(k)\equiv  
\sqrt{\Delta^2_g(k)/\Delta^2_{\rm dm}(k)}$ and  the gas-dark matter
cross correlation coefficient $r\equiv \Delta^2_{\rm dg}(k)/\sqrt{\Delta^2_{\rm
dm}(k)\Delta^2_g(k)}$. We expects $b_g(k)<1$ at very nonlinear
scales. This behavior is observed in our simulations
(Fig. \ref{fig:r}). Poorer resolution of gas with respect to dark
matter means gas is smoother than dark matter, so  phenomenologically,
one can treat the gas density field as a smoothing of the underlying
dark matter density field:
\begin{equation}
\delta_g({\bf x})=\int \delta_{\rm dm}({\bf x}^{'}) W_g({\bf x-x}^{'}) d^3 x^{'}.
\end{equation} 
In this model, gas perfectly
correlates with dark matter and  $r\equiv 1$. In our simulation, we
find that at $\Delta^2_{\rm dm}\la 200$, this is the case
(Fig. \ref{fig:r}).  One  can model $W_g(k)=\exp[-k^2/k_g^2]$. An ideal 
simulation should have a $k_g$ such that $\Delta^2_g(k_g) \gg 1$.  In
simulations, $k_g(z)$ should increase with $z$ since for higher z, the
nonlinear scale is smaller (e.g. fig. \ref{fig:r}). The simulated
$k_g$ can be modeled by $k_g(z)=5(1+z)^2$ h/Mpc, which is roughly
consistent with the  simulated gas power spectrum, and reproduces the
simulation results (Fig. \ref{fig:wmapTz}).  This 
agreement implies that for WMAP cosmology, the simulation resolution causes the
simulated $\bar{T}_g$ to lose power at low redshift.

\begin{figure}
\epsfxsize=9cm
\epsffile{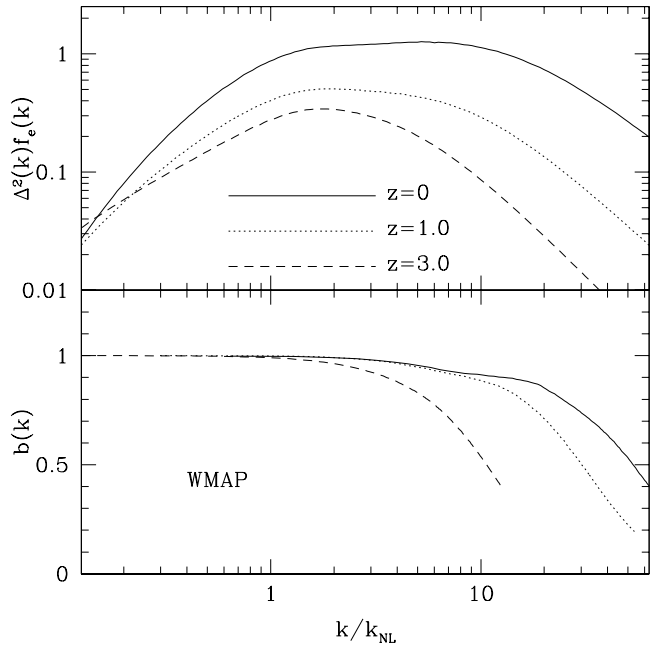}
\caption{The requirement of $\bar{T}_g$ on simulation resolution. The
top panel plots the integrand of Eq. \ref{eqn:meanT}, namely, the
contribution of different scales to $\bar{T}_g$. The nonlinear power
spectra are calculated by the \citet{Smith03} code. At $z\ga 1$, only $k\sim
k_{\rm NL}$ is required to resolve. But at $z=0$, $k\ga 10 k_{\rm NL}$
is required to resolve. The bottom panel is the gas bias in our
simulation. Its deviation from unity is a measure of the simulation
resolution. We find that, at $z\ga 1$, WMAP simulation meets the
resolution requirement. But at $z=0$, the resolution requirement
($k\sim 10 k_{\rm NL}$) is beyond simulation ability. Since simulated power
spectrum loses power at $10\%$ level at $k\sim 10 k_{\rm NL}$, we
expect the simulated $\bar{T}_g$ to lose power at $10\%$ level, as
predicted in Fig. \ref{fig:wmapTz}.  
\label{fig:res}} 
\end{figure} 

This conclusion seems counter-intuitive.  Since gas temperature arises
from thermalization at nonlinear scales, which are better resolved in
lower redshift simulations (left panels of fig. \ref{fig:r} and bottom
panel of fig. \ref{fig:res}), we may expect less severe resolution
problem for simulated $\bar{T}_g$  at lower redshifts. For self
similar simulations, this is true since $k_{\rm NL}$ is the only
relevant scale. But $\Lambda$CDM cosmology breaks
self similar condition and makes the nonlinear scale $k_{\rm NL}$ not the
only relevant  parameter to determine $\bar{T}_g$. We show how this
running power index causes more severe resolution problem for simulated
$\bar{T}_g$. 

The relative contribution from different scale $k$ to $\bar{T}_g$
relies on the slope of the power spectrum. The larger the power index
at $k>k_{\rm NL}$ is, the larger the relative contribution to
$\bar{T}_g$ is and the higher the requirement of  spacial resolution in unit of
$k_{\rm NL}$ to simulate $\bar{T}_g$ is. For WMAP cosmology, $k_{\rm
NL}$ keeps decreasing toward low z. The effective power index at
$k=k_{\rm NL}$ keeps increasing and the relative contribution from
$k>k_{\rm NL}$ 
keeps increasing. Such behavior requires a stronger spacial resolution in unit
of $k_{\rm  NL}$. In order for the simulated $\bar{T}_g$ not to lose
power, at $z=1$, simulation must be able to resolve $k\la 2k_{\rm
NL}$, but at $z=0$,  the requirement is $k\la 10 k_{\rm NL}$
(Fig. \ref{fig:res}).  As we see from the bottom panel of
Fig. \ref{fig:res}, WMAP simulation meets this requirement at $z=1$ but
fails at $z=0$. So, we conclude that, for CDM simulation with
reasonable large simulation box ($\ga 50$ Mpc/h), the simulation
resolution is the dominant simulation limitation to simulate
$\bar{T}_g$.

{\bf 
In the above discussion, we have assumed that in the relevant $k$
range, $\Delta^2_g=\Delta^2_{\rm dm}=\Delta^2_{\rm dg}$. Though at highly
nonlinear regime, the simulated $\Delta^2_g$ does lose power with respect
to $\Delta^2_{\rm dm}$ (Fig. \ref{fig:r}), this is likely caused
by resolution effect since when increasing resolution,
$b^2=\Delta^2_g/\Delta^2_{\rm dm}$ keeps increasing toward unity (Fig.
9).  Since we do not have higher resolution simulations to test if $b$
will reach unity in these nonlinear regimes, this question still
remains open. But one can estimate it using the halo model. One expects that 
the significant deviation from $b=1$ could only happen where gas pressure is
comparable to gravity, namely, in each halos. Observationally, intracluster gas
develops a constant density core while dark matter
may develop a density  cusp toward the center. It is not clear whether
gas pressure alone could generate such gas core or other mechanism
such as feedback is required.  But even the deviation in the
gas and dark matter  distribution presents in adiabatic simulations
(with infinite resolution), it should only happen at very small scale
$r\la r_c\sim 0.1$ Mpc/h, or $k\ga 10 h/$Mpc. Thus we expect that
$b=1$ is a good approximation at scales $k\la 10 h$/Mpc. Since the
dominant contribution to $\bar{T}_g$ for WMAP cosmology comes from
$k\ll 10 h$/Mpc, we do not expect the possible real deviation of $b$
to be responsible for the discrepancy between predicted and simulated
$\bar{T}_g$. We thus neglect this possible effect. }  

\begin{figure}
\epsfxsize=9cm
\epsffile{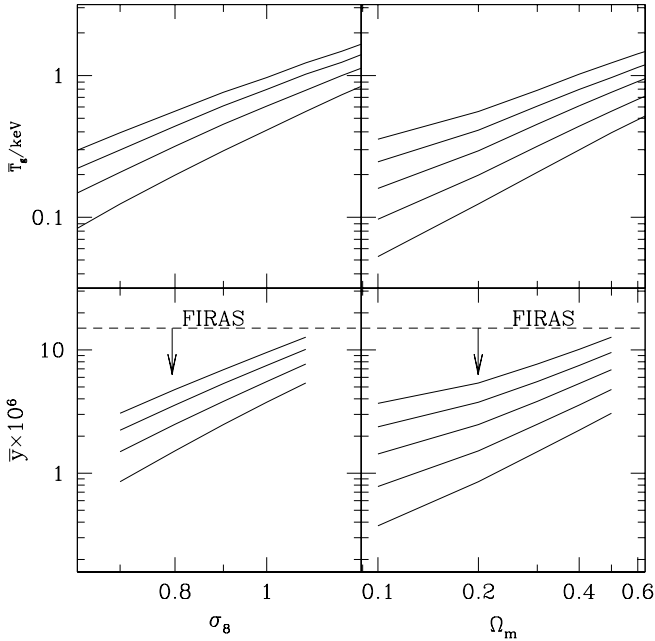}
\caption{The predicted $\bar{T}_g$ and $\bar{y}$ as a function of
$\sigma_8$ and $\Omega_m$. For the left panels,
from top down, lines correspond to $\Omega_m=0.5,0.4,0.3,0.2$. For the
right panels, from top down, lines correspond to
$\sigma_8=1.1,1.0,0.9,0.8,0.7$. \label{fig:T_cosmology}}
\end{figure}

In summary, our choice of $r_e$, the only free parameter in our model
passes the tests of all simulations. Thus our model has no free
parameter,  is free of simulation artifacts  and is able to predict the
real $\bar{T}_g$ to several percent accuracy. 

{\bf It is interesting to compare these results with the halo model
prediction. For the
self similar case, the same scaling relation
(Eq. \ref{eqn:selfsimilarTz}) is predicted. The scaling relation  does not
depend on the specific form of the halo mass function, either
Press-Schechter formalism or the Jenkins fitting formula. It only
relies on (1) the virial theorem $T_g\propto m^{2}/r_{\rm vir}$ and (2)
the general form of halo mass function $dn/dm\propto m^{-2}
f(\nu)|d\ln\nu/d\ln m|$. The detailed study \citep{Refregier02} shows
that the halo model prediction is consistent with
simulations. Surprisingly, the halo model also tends to predict a
higher $\bar{T}_g$ at low $z$ than simulations and strongly implies
the same resolution effect. The continuum model and the
halo model rely on different assumptions and require different input.
Though both models are based on virial theorem, the continuum model
treats gas as continuum field and the halo model treats gas as
distributed in discrete halos. The continuum model requires the
nonlinear density power spectrum, while the halo model requires the halo
mass function and the halo $m-T$ relation.  The consistency in the
predictions of two distinctive models gives us confidence that the
predicted $\bar{T}_g$ is reliable and simulations artifacts are
suitably handled. }  

For the WMAP cosmology, we predict $\bar{T}_g(z=0)=0.32$
keV. $\bar{T}_g$ is sensitive to $\sigma_8$ and $\Omega_m$. For self
similar cosmology, $\bar{T}_g\propto \sigma_8^{-(n-1)/(n+3)}$. For $\Lambda$CDM, the actual  dependences of $\bar{T}_g$ on
$\sigma_8$ and $\Omega_m$ (fig. \ref{fig:T_cosmology}) is 
complicated due to the running index of the density power
spectrum. In our interested $\sigma_8$ and $\Omega_m$
range, the effective power index $n_{\rm eff}$ is $-2\la
n_{\rm eff}\la -1$. So  $\bar{T}_g\propto
\sigma_8^{\alpha}\Omega_m^{\beta}$ with $\alpha\sim
-(n_{\rm eff}-1)/(n_{\rm eff}+3)\sim  
1$-$3$ and $\beta\sim 1$.  A smaller $\sigma_8$ results in a larger
$k_{\rm NL}$ and thus a smaller effective power index $n_{\rm eff}$. So
$\alpha$ is larger. The deviation of $\beta$ from unity comes from the
dependence of $n_{\rm eff}$ on $\Omega_m$ since CDM transfer function
depends on the combination $q\simeq k/\Omega_m$.  Around the WMAP
cosmology $\Omega_m=0.268$ and $\sigma_8=0.84$, $T_g$ can be fitted as
\begin{equation}
T_g=0.32
(\sigma_8/0.84)^{3.05-0.15\Omega_m}(\Omega_m/0.268)^{1.28-0.2\sigma_8}\
{\rm keV}.
\end{equation}

\section{The SZ mean $y$ parameter}
\begin{figure}
\epsfxsize=9cm
\epsffile{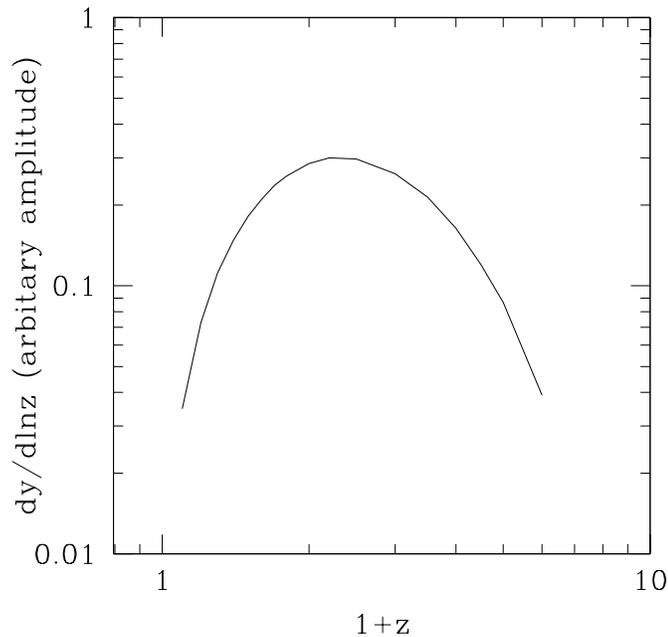}
\caption{The integrand of $\bar{y}$, as defined by $dy/d\ln z$ for the
WMAP cosmology. The dominant contribution to $\bar{y}$ comes from
$z\sim 1$. \label{fig:yz}}
\end{figure}

The SZ mean $y$ parameter is calculated using Eq. \ref{eqn:y} and the
result is shown in Fig. \ref{fig:T_cosmology}. $\bar{y}$ is generally
$\sim 10^{-6}$. For the WMAP cosmology,
$\bar{y}=2.6\times 10^{-6}$ and the mean temperature decrement at
Rayleigh-Jeans regime is $14 \mu$K. The dominant contribution comes
from $z\simeq 1$ (Fig. \ref{fig:yz}). Since at $z\ga 1/2$, $d\chi/dz\propto 1/
\sqrt{\Omega_m}$, one may expect $\bar{y}\propto
\bar{T}_g/\sqrt{\Omega_m} \propto \Omega_m^{\sim 0.5}$. But since in a
higher matter density universe, the density field evolves faster and
thus $\bar{T}_g$ drops faster with increasing $z$, the $\bar{y}$
dependence on $\Omega_m$ is stronger than  $\Omega_m^{\sim 0.5}$. Indeed, we find $\bar{y} \propto \Omega_m^{\sim 1}$.   Around the WMAP
cosmology $\Omega_m=0.268$ and $\sigma_8=0.84$, $\bar{y}$ can be fitted as
\begin{equation}
\bar{y}=2.6\times 10^{-6}
(\sigma_8/0.84)^{4.1-2\Omega_m}(\Omega_m/0.268)^{1.28-0.2\sigma_8}.
\end{equation}

Though our model is able to predict the Compton $\bar{y}$ in an
adiabatically evolving  universe to several percent  accuracy, it does
not include any non-gravitational thermal processes, which introduce
non-negligible effect to $\bar{y}$. Photo-ionization
contributes  $\bar{y}_{\rm photonion}\sim \tau 10^4 {\rm
K}/m_ec^2\sim  3\times 10^{-7}$, or $\sim 10\%$ of the adiabatic IGM
$\bar{y}$.  Though feedback, preheating,
radiative cooling may decrease the SZ power spectrum by a factor of
$2$ \citep{daSilva01,Lin02,White02,Zhang03}, they  only 
affects $\bar{y}$ at $10\%$ level (e.g. \citet{White02}). This is
straightforward to  
understand. Once hydrostatic equilibrium is reached, the gas pressure is 
always determined by the gravitational potential, which is mainly set
by dark matter distribution and is only weakly affected by these
thermal processes.  Feedback and preheating do not change the total
amount of gas. While radiative cooling turns some gas into bound
objects, such mass loss is minor since bounds objects in galaxies only
accounts for $\la 10\%$ baryons. Thus, the change of the total gas
thermal energy due to these processes is minor. These processes
change $\bar{y}$ mainly  during the stages 
of expansion in the case of feedback and preheating and infall in the
case of radiative cooling. Such stages are either in semi hydrostatic
equilibrium (feedback and radiative cooling) or last relatively short
time (mild preheating), thus their effects to $\bar{y}$ is not
significant. Cluster magnetic field also only has $\sim 10\%$ effect
to $\bar{y}$ \citep{Zhang03b}. 

WMAP measured a high Thomson optical depth to 
the last scatter surface and implies an early reionization epoch
caused by first stars. At high $z$, CMB density is high and is able to
convert a considerable fraction of first star supernova explosion
thermal energy through the efficient Compton cooling.
Such first star  contribution to $\bar{y}$ is $\sim$ few $10^{-6}$ and
comparable to low redshift IGM $\bar{y}$
\citep{Oh03}. 

These processes have distinctive signatures to $\bar{y}$ and the SZ
power spectrum, respectively. For first stars,  since at high $z\sim   
10$, density fluctuations are small, the relative contribution of
first stars to the SZ fluctuation  is much 
smaller than that of to $\bar{y}$. As estimated by \citet{Oh03}, even
if first star $\bar{y}$ is 
larger than that of low redshift IGM, its contribution to the SZ power
spectrum can 
still be an order of magnitude smaller than that of low redshift IGM
(Fig. 1, \citet{Oh03}). The case of photon-ionization is similar, but that
of feedback, preheating, radiative cooling and magnetic field is
opposite. Thus, combining the $\bar{y}$ and the power spectrum
measurement helps to separate these contributions more
unambiguously. For example, if $\bar{y}$ can be measured to $\sim
10\%$ accuracy, the first star contribution can be constrained with a
statistical uncertainty $\sim 0.4\times 10^{-6}$ and systematic
underestimation of $\sim 0.5)\times 10^{-6}$ caused by feedback, etc.

Unfortunately, the direct precision measurement of absolute $\bar{y}$
is difficult. Currently, the best measurement, $\bar{y}<1.5\times
10^{-5}$ is given by  the COBE/FIRAS measurement
\citep{Fixsen96}. This result has already been able to rule out
combinations of high $\sigma_8\ga 1.1$ and high  $\Omega_m\ga
0.5$. This constrain is quite weak. But considering the contributions
from  non-gravitational processes could make it stronger.  
Nonetheless, $\bar{y}$  may be measured to a higher accuracy in the
future and/or 
inferred from  new statistics and helps to independently constrain
$\sigma_8$ and $\Omega_m$.

\section{Summary}
The mean  gas density weighted temperature $\bar{T}_g$ and the mean SZ Compton $y$ parameter are  the lowest order SZ
statistics. Their precision modeling stands as the first step toward the
precision understanding of the IGM SZ effect and may provide useful
clues for modeling of higher order statistics. The two ways of the SZ
modeling, analytical models and hydro  
simulations both have their own limitations. It is essential to
quantify simulations limitations and test analytical models against
corrected simulations. The convergence of $\bar{T}_g$ stands as the
lowest requirement for simulations to reliably predict the SZ
effect. 

We ran $n=-1,-2$ self similar $512^3$ hydro 
simulations to quantify simulation limitations utilizing their self
similar scaling relation. We also ran a high resolution $1024^3$ cell, 100
Mpc/h box size hydro simulation adopting WMAP cosmology. We find that
the simulated  $p(>T)$, the fraction of mass with temperature bigger
than $T$,  shows a good convergence for all our simulations, except at
both tails. This convergence suggests that $\bar{T}_g$ is well
simulated. Our
continuum field model is then tested against these simulations. It
passed all tests and we believe that its prediction for $\bar{y}$ is
accurate to several percent. 

Various simulation limitations such as limited box size and limited
resolution can affect the simulated $\bar{T}_g$. Generally, for a $\Lambda$CDM
simulation, the nonlinear scale is much smaller than the box size,
thus the box size effect is negligible and the resolution effect is
the dominant cause of simulation artifacts in $\bar{T}_g$. Due to
curvature in the CDM power spectrum, the temperature is more
difficult to model accurately in simulations at fixed comoving resolution.
We found that, at $z=0$, due to the
simulation resolution, gas power spectrum loses power at small scales
with respect to dark matter power spectrum. This behavior causes the
simulated gas density weighted temperature $\bar{T}_g$ to be $\sim 20\%$
underestimated. But this resolution effect becomes negligible quickly
toward higher redshift. At $z\ga 0.5$, simulated $\bar{T}_g$ is quite
accurate. Since the dominant contribution to $\bar{y}$ comes from
$z\sim 1$, our simulation prediction of $\bar{y}$ is reliable to $\sim
10\%$ level. Furthermore, our analytical model is able to correct this
simulation artifacts and predicts $\bar{y}$ with several percent
accuracy. For a flat, low matter density  $\Lambda$CDM universe,
$\bar{y}=2.6\times 10^{-6}
(\sigma_8/0.84)^{4.1-2\Omega_m}(\Omega_m/0.268)^{1.28-0.2\sigma_8}$. 
The current upper limit of $y<1.5\times 10^{-5}$ measured by FIRAS has
already ruled out combinations of high $\sigma_8\ga 1.1$ and high
$\Omega_m\ga 0.5$.

Our simulations confirms previously found $T\propto \rho$ relation in
a large region of $\rho$-$T$ plane.  Simulation resolution may be
partly responsible for this relation. But the results of self similar
simulations show robust convergence of this relation and can not be
explained by simulation resolution artifacts.  This issue deserves a
further investigation. We 
also found that,  the simulated $p(>T,>\delta)$, does not converge. At
high density regions,  it is 
caused by density resolution limitation while at low density regions, it is
caused by failure of capturing shocks. Though this numerical limitation has
only minor effect on $\bar{T}_g$, it may affect the gas pressure power
spectrum a lot. This issue will be addressed in a companion paper. 

\section*{Acknowledgments}
We thank the anonymous referee for many helpful comments, especially
the comparison with the halo model. 
Computations were performed on the CITA Pscinet computers funded by
the Canada Foundation for Innovation.  P.J. Zhang is  
supported by the DOE and the NASA grant NAG 5-10842 at Fermilab.

\end{document}